\documentstyle[aps,multicol,epsf]{revtex}

\begin{document}

\title{Towards Granular Hydrodynamics in Two-Dimensions}
\author{E.\ L.\ Grossman$^1$\thanks{Address e-mail to:
grossman@cs.uchicago.edu}, Tong Zhou$^1$, and E.\ Ben-Naim$^{1,2}$}
\address{$^1$The James Franck Institute, The University of Chicago, 
Chicago, IL 60637}
\address{$^2$Theoretical Division and Center for Nonlinear Studies, 
Los Alamos National Laboratory, Los Alamos, NM 87545}
\maketitle
\begin{abstract}
We study steady-state properties of inelastic gases in two-dimensions
in the presence of an energy source. We generalize previous
hydrodynamic treatments to situations where high and low density
regions coexist. The theoretical predictions compare well with
numerical simulations in the nearly elastic limit. It is also seen
that the system can achieve a nonequilibrium steady-state with
asymmetric velocity distributions, and we
discuss the conditions under which such situations occur.

\vspace{0.1in}
\end{abstract}
\begin{multicols}{2}
\section{Introduction}
Granular materials such as sand and powders have generated much
interest of late. Such an ensemble of particles with macroscopic
size is challenging since it may behave as a solid, a liquid or 
a gas.  Size separation, pattern formation, avalanches, compaction 
and convection are just a few examples of the wide array of observed 
phenomena\cite{JNB}. 

Flow underlies most of these phenomena and therefore, theoretical 
studies so far focus on formulating a hydrodynamical description 
appropriate to sand\cite{Haff,JenkRich}.  These theories, stemming from
the Boltzmann Equation, depend on the assumption of ``molecular 
chaos'', {\it i.e.}, the assumption that no 
interparticle correlations exist.  This assumption is far from obvious.
As a dissipative dynamical system, a 
granular system has attractors in its phase space, which may cause
correlations between particles. Under certain conditions, 
these attractors lead to a singularity, inelastic collapse, which can 
not be explained by hydrodynamics. In one dimension, these 
attractors are so strong that hydrodynamics breaks down for the 
entire parameter space\cite{DLK,GR}---in a confined geometry, 
all particles,
save one, form a practically stationary clump against the 
elastic wall, while the remaining particle moves rapidly back and 
forth between the clump and the heated wall.  Such a state clearly 
violates partition of energy.  

\begin{figure}
\narrowtext
\epsfxsize=\hsize
\vspace{-.2in}
\epsfbox{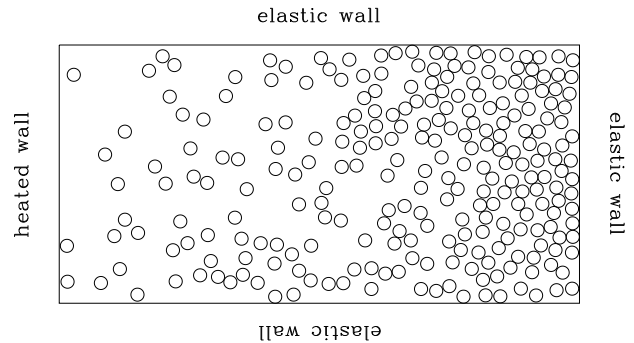}
\vspace{-.2in}
\caption{The system of interest.  The length of the
system is $L$, the width is $W$.  The heated wall is at $x=0$;
the elastic walls are at $x=L$, $y=0$, and $y=W$.}
\label{fig:syspict}
\end{figure}

In this study, we investigate the corresponding situation in
two-dimensions\cite{Mad,EsiPoe}. We consider inelastic hard 
spheres in a box where one
wall is kept at a fixed temperature and the other three are reflecting
(see Fig.~\ref{fig:syspict}). Energy input at the heated wall balances
the dissipation due to inter-particle collisions and the system can
achieve a steady state. Unlike the one-dimensional situation, the
density and the temperature profiles are smooth functions of the
distance from the heated wall.  In the steady-state, the momentum
balance equations imply that the
pressure is constant throughout the system. Particles move 
faster close to the energy source and more slowly deeper inside the system 
due to energy loss. Thus the density is greater farther away from
the wall to maintain a constant pressure.

Density variations may be a consequence of the above mechanism or an 
effect of the intrinsic attractors in the system which build up 
correlations among particles.  Assuming that the particles are 
not coherent in the quasi-elastic limit, we derive a differential 
equation that describes the density variation throughout the system.
Heuristic expressions for the
mean-free path, the equation of state, and the thermal conductivity
are incorporated into the energy flux balance equation to obtain a
closed equation describing the steady-state density profiles.  The theory
compares well with simulational results over a wide range of densities in
the quasi-elastic limit.

Our theory is appropriate for a steady-state which is very close to an 
equilibrium state.  We expect this to be applicable only in the nearly 
elastic limit. We argue that in the
complementary situation of stronger inelasticity, hydrodynamics may still
be relevant. To analyze the nature of the steady-state, we study
velocity distributions throughout the system.  We observe 
that the velocity distributions exhibit scaling. This observation
is then used to obtain a qualitative description of the behavior of 
the system.

\section{The quasi-elastic limit}

Granular materials are different from ordinary fluids or gases in that 
the diameter of a particle may be comparable to the collisional mean 
free path.  To formulate a hydrodynamic theory, it is
necessary to describe how the mean free path, the pressure, and the
thermal diffusivity depend on the grain diameter, the number density,
and the temperature. In this section, we obtain heuristic expressions
for these quantities.

Let us denote the number density of grains by $\rho$, the mean free
path by $l$, and the grain diameter by $d$. Without loss of
generality, we set the particle mass and the Boltzmann constant to
unity: $m=k_B=1$.  The granular temperature can be defined as the
average kinetic energy per 
particle: $T\equiv \frac{1}{2} \langle v^2 \rangle$.  
This quantity is well defined for an equilibrium state in
which the particles have a symmetric velocity distribution. In such a
case, $\bar{v} = \sqrt{2T}$, can be used as an approximate value for the
average grain speed.  In our system, quantities such as temperature, 
density, etc., are position dependent, and there is not a global 
equilibrium.  However, in the quasi-elastic limit, we expect the 
system to be very close to local thermal equilibrium.

Let $x$ be the distance from the heated wall.  In
the steady state, all quantities vary only in this transverse
direction. The energy balance equation can therefore be written
\begin{equation}
\frac{dq}{dx}+I=0,
\label{eqn:enerbal}
\end{equation} 
where $q$ is the energy flux in the $x$ direction. The sink 
term $I$ accounts for the
energy lost per unit area per unit time due to inelastic collisions.
The energy flux is induced by a temperature gradient, $q=-\kappa dT/dx$, 
where the coefficient of thermal
diffusivity, $\kappa$, is positive. Consequently, we find  
\begin{equation}
\frac{d}{dx}\left(\kappa\frac{dT}{dx}\right)=I.
\label{eq:m2}
\end{equation}

\subsection{The Energy Sink}

Since collisions between grains are inelastic, kinetic energy is
continually transferred into heat.  For simplicity we neglect rotation
and thus, the degree of inelasticity  can be parameterized by $0\le r \le 1$.
When two particles collide, their tangential velocities are
unchanged, while the relative normal velocity is decreased by a factor
of $-r$, where the negative sign merely indicates that they move apart after
a collision.  Using momentum conservation one can write the final
velocities (indicated by primes) in terms of the initial velocities 
\begin{equation}
\left( \begin{array}{c} v_{1n}' \\ v_{2n}' \end{array} \right) = 
\frac{1}{2} \left( \begin{array}{cc} 1-r & 1+r \\ 1+r & 1-r \end{array} \right)
\left( \begin{array}{c} v_{1n} \\ v_{2n} \end{array} \right).
\end{equation}
In the above equation, the subscript $n$ denotes the velocity component
along the line connecting the centers of particles 1 and 2.
The energy lost in each collision is therefore
$$ \Delta E =-\frac{1}{4}(1-r^2)(v_{1n}-v_{2n})^2. $$
In this study, we focus on the  quasi-elastic limit, i.e.\ $1-r \ll 1$.
Physically, this limit is relevant to hard particles
such as glass or steel beads.

Using the above expression for the energy dissipated in each collision,
one can estimate the
sink term, $I$, the mean energy lost per unit area per unit time.  Consider a
particle moving with speed $\bar{v}$. During each collision,
it loses, on average, energy proportional to $(1-r^2)\bar{v}^2$.  In unit
time, it collides roughly $\bar{v}/l$ times. In unit area, there are
$\rho$ particles, and consequently 
$$I \propto
(1-r^2)\bar{v}^2\frac{\bar{v}}{l}\rho \propto (1-r^2)\rho T^{3/2}/l. $$

\subsection{The Coefficient of Thermal Diffusivity}

Suppose there is a temperature gradient in the $x$-direction.  As a
result, there will be an energy flux along this axis.  To calculate
this flux, let us consider the number of particles crossing a line
perpendicular to this direction in a time interval
$\Delta t$. We define ``crossing the line'' as having any part of the
particle over the line during this time interval.  To cross the line
within $\Delta t$, a particle with speed $\bar{v}$ must have its rightmost
point $\bar{v}\Delta t$ or closer to the line.  Thus, only grains in an area
($d + \bar{v}\Delta t$) to the left of the line can pass the line from the left
in $\Delta t$.  In fact, only one half of these particles are moving to the
right, so the number of particles crossing the line per unit
cross-sectional length 
in $\Delta t$ is $\frac{1}{2}\rho (d + \bar{v}\Delta t)$.  
In a steady state, the number of
crossing events from each side must balance.  Any energy flux is
due to the fact that particles coming from the right are at a
different temperature ($T_R$) than those from the left ($T_L$) and
thus, $q\Delta t = \frac{1}{2}\rho (d + \bar{v}\Delta t)(T_L-T_R)$.  
Consider two
grains on opposite sides of the line.  Their centers are approximately
a distance $(l+d)$ apart, so the temperature difference is roughly
$(l+d)dT/dx$.  The coefficient of thermal conductivity is therefore
$\kappa = -q/(dT/dx) = \frac{1}{2}\rho (l+d)(d+\bar{v}\Delta t)/\Delta t$. 
A natural
choice for $\Delta t$ is the typical collision time $l/\bar{v}$. This 
choice is small enough to avoid multicounting and is sufficiently 
large to ensure that heat transfer does occur. Our heuristic picture
therefore estimates the thermal diffusivity by
$$\kappa \propto \frac{\rho (l+d)^2 \sqrt{T}}{l}.$$
This is a rough approximation; the actual prefactors depend on
the velocity distribution of the grains.  Therefore, we generally
assume
$$\kappa \propto \frac{\rho (\alpha l+ d)^2 \sqrt{T}}{l}.$$

The energy balance equation~(\ref{eq:m2}) takes the form
\begin{equation}
\frac{d}{dx}\left[\frac{(\alpha l + d)^2}{l}\rho\sqrt{T}\frac{dT}{dx}\right]
= (1-r^2)\frac{\rho T^\frac{3}{2}}{\gamma l}
\label{eq:m3}
\end{equation}
where $\gamma$ is the ratio of prefactors in the expressions 
for $I$ and $\kappa$.  In the following subsections, we discuss how to 
obtain the dimensionless
coefficients $\alpha$ and $\gamma$ self-consistently. 
To proceed, it is necessary to relate $\rho$ and $T$ through the 
equation of state. Additionally, the mean free path $l$ must be
expressed in terms of $\rho$.

\subsection{The Equation of State}

For the system to be in a steady state, the pressure must be constant
throughout. The equation of state relates the pressure and the
temperature to the
density.  For example, in the low density limit, the ideal gas law
holds, $PV = Nk_BT$. Using $k_B=1$ and $\rho = N/V$, one has 
\begin{equation}
P = \rho T.
\label{eqn:pressld}
\end{equation}
On the other hand, in the high density limit, the mean free path, which is
simply the interparticle spacing, is much less than the particle diameter
($l \ll d$). 
Denoting the close-packing density by $\rho_c$, one finds 
\begin{equation}
\frac{\rho}{\rho_c}=\frac{d^2}{(d+l)^2}\approx1-\frac{2l}{d}.
\label{eqn:denl}
\end{equation} 
In this limit, the center of a grain is confined to an area of
the order of $l^2$, so the entropy per particle, $\cal S$, 
equals $\ln (l^2)$ plus a function of temperature.  From
Eq.~(\ref{eqn:denl}), $\cal S$ depends on density only 
through the term $2\ln (\rho_c - \rho)$.  Using Maxwell's relation,
$$ \left(\frac{\partial P}{\partial T}\right)_{1/\rho}
= \left(\frac{\partial {\cal S}}{\partial\frac{1}{\rho}}\right)_T
= \frac{2\rho^2}{\rho_c-\rho},$$
we obtain the pressure in the limit $\rho \rightarrow \rho_c$,
\begin{equation}
P = \frac{2\rho^2T}{\rho_c-\rho}.
\label{eqn:presshd}
\end{equation}
We therefore propose the following interpolation formula for 
the pressure
\begin{equation}
P = \rho T \frac{\rho_c + \rho}{\rho_c-\rho}.
\label{eqn:pressgen}
\end{equation}
Indeed, in the limits of high and low density this expression reduces
to (\ref{eqn:pressld}) and (\ref{eqn:presshd}), respectively.  It
is useful to compare this expression with the van der Waal's
equation of state which takes into account long range attraction and
hard core repulsion \cite{hay}.  For an inelastic gas of hard spheres,
there are no long range
forces, and the van der Waal's pressure for a two-dimensional gas reads
$P_{vdW} = \rho\rho_c T/(\rho_c-2\rho)$ for $\rho \ll \rho_c$.  In the low
density limit the pressure given by Eq.~(\ref{eqn:pressgen})
agrees with the van der Waal's expression to second order in
$\rho/\rho_c$.  Furthermore, one can also compare (\ref{eqn:pressgen}) to
Tonks' series expansion~\cite{Tonks} for the 
pressure of a two-dimensional gas
of hard spheres, $P_{Tonks} = \rho T[1 + 1.814(\rho/\rho_c) +
2.573(\rho/\rho_c)^2]/[ 1 - 1.307(\rho/\rho_c)^3 + 0.307(\rho/\rho_c)^4]$, 
valid for all densities.  Over the entire density range, the
two expressions differ by less than 1.3\%.  
In contrast, we found that the van der Waal's
formula is inadequate for describing the high density limit.  Hence, we
use the interpolation formula (\ref{eqn:pressgen}) for the pressure.

\subsection{The Mean Free Path}

The mean free path can be expressed in terms of the density and 
the diameter. In the low density limit one has 
$$l=\frac{1}{\sqrt{8}\rho d},$$
while in the high density limit, Eq.~(\ref{eqn:denl}) gives 
$$l=\frac{\rho_c-\rho}{2\rho_c} d.$$
Again, we use these high and low density limits to interpolate 
a general expression for the mean free path.
Using the 2D close-packing value $\rho_c = 2/\sqrt{3}d^2$, we find 
\begin{equation}
l \approx \frac{1}{\sqrt{8}\rho d}\frac{\rho_c-\rho}{\rho_c-a\rho},
\label{eqn:lgen}
\end{equation}
where $a=1-\sqrt{3/8}$.

\subsection{The density equation}

Eqs.~(\ref{eqn:pressgen}) and~(\ref{eqn:lgen}) express
the temperature and the mean free path in terms of the density.  
Substituting these expressions into Eq.~(\ref{eq:m3}) yields a
second order differential equation for $\rho$.  
Using for convenience the variable $z \equiv \rho_c/\rho$, we have
\begin{eqnarray}
\frac{d}{dx}
\left[\frac{(z^2+2z-1)
\left(\alpha z(z-1)+\sqrt{\frac{32}{3}}(z-a)\right)^2}
{(z-a)(z-1)^{1/2}z^{3/2}(z+1)^{5/2}}
\frac{dz}{dx}\right] =  \nonumber\\
& & \hspace{-6cm}\frac{32}{3d^2}
\frac{1-r^2}{\gamma}
\frac{z-a}{(z+1)^{3/2}}
\sqrt{\frac{z-1}{z}}.
\label{eqn:zgen}
\end{eqnarray}
This equation is complemented by the boundary conditions
\begin{equation}
\frac{dz}{dx}\Big|_{x=L}=0\qquad\mbox{and}\qquad W\int_0^L\frac{\rho_c}{z}dx=N,
\label{eqn:bcnorm}
\end{equation}
where $W$ is the width of the system and $N$ is the total 
number of particles.  The latter condition merely reflects
conservation of particles.  The former condition is 
a consequence of the fact that the temperature gradient
vanishes at the elastic wall.

In principle, $\gamma$ and $\alpha$ are two dimensionless factors
which can be calculated exactly from the velocity distribution.
However, this distribution is poorly understood.  
Nevertheless, it
is still possible to estimate these prefactors by comparing the
theoretical predictions of Eq.~(\ref{eqn:bcnorm}) with numerical
simulations in the limit of high and low densities. 

In both of these extreme cases the governing equation~(\ref{eqn:zgen})
can be solved analytically.  In the high density 
limit, $(\rho_c-\rho) \rightarrow 0$,
or equivalently, $z-1 \ll 1$, Eq.~(\ref{eqn:zgen}) reduces 
to $\frac{d^2}{dx^2}\sqrt{z-1}=\frac{1-r^2}{2\gamma d^2}\sqrt{z-1}$.  It is
convenient to write this equation in terms of the temperature, $T(x)$.
From~(\ref{eqn:pressgen}), $T(x) \propto z-1$ when $z-1 \ll 1$, and 
the temperature obeys
$$\frac{d^2}{dx^2}\sqrt{T(x)} = \frac{1}{4\xi^2}\sqrt{T(x)}$$
with $\xi \equiv d\sqrt{\gamma/2(1-r^2)}$.  
Solving this equation subject to the boundary condition
$dT/dx=0$ at $x=L$ gives the temperature profile 
$$T(x)=T(L)\cosh^2\left(\frac{L-x}{2\xi}\right).$$
Far from the elastic wall, $L-x \gg \xi$, the temperature decays
exponentially in agreement with Haff's calculation \cite{Haff} 
\begin{equation}
T(x)=T(0)e^{-x/\xi}.
\label{eqn:hdtemp}
\end{equation}
Eq.~(\ref{eqn:lgen}) shows that in the high density 
limit, $l \propto z-1 \propto T$.
Both the temperature and the mean free path decay exponentially
with the distance from the heated wall.  The decay length $\xi$ is
much larger than the mean free path.  For the continuum description
to be valid, $\xi$ must also be much larger than the diameter of a
particle, $d$, i.e.,
\begin{equation}
\sqrt{1-r^2}\ll \sqrt{\frac{\gamma}{2}}.
\label{eq:valid1}
\end{equation}
Since the prefactor $\gamma$ is of order unity, we learn that the 
continuum theory is valid only in the quasi-elastic 
limit, $\sqrt{1-r} \ll 1$.  (Note that the restriction $\sqrt{1-r} \ll 1$
is stricter than $1-r \ll 1$.)

To test the theoretical predictions, we performed an event-driven
simulation of the system\cite{goldzan}.  The heated wall is implemented
in such a way that for $r=1$, particles have a Boltzmann velocity 
distribution with an average temperature equal to
one.  Specifically, any particle 
that collides with the wall at $x=0$ is ejected with a positive $v_x$
drawn from the probability distribution $P(v_x) = v_x \exp(-v_x^2/2)$
and a $v_y$ drawn from $P(v_y) = \exp(-v_y^2/2)/\sqrt{2\pi}$\cite{GR}.
However,  the behavior of particles in the bulk of the system is independent 
of the details of the boundary condition (see Fig.~\ref{fig:vxscal} below).
This is not surprising, but rather a necessity for a thermodynamic theory 
to be valid. 

\begin{figure}
\narrowtext
\epsfxsize=\hsize
\vspace{-.1in}
\epsfbox{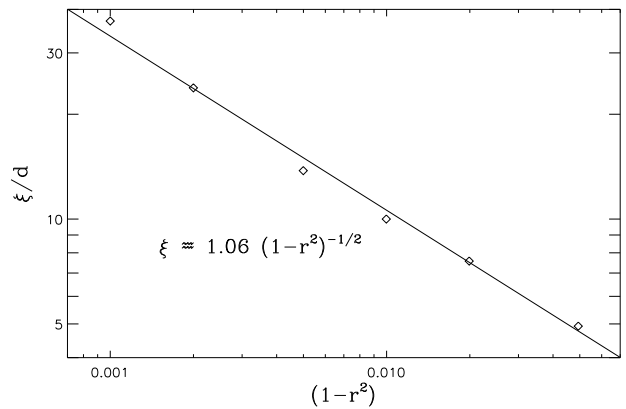}
\vspace{-.1in}
\caption{The behavior of the decay length $\xi$ in units of particle
diameters as a function
of $1-r^2$ in the high density limit.  The solid 
line is $1.06 (1-r^2)^{-1/2}$, and the diamonds are values obtained 
from simulations with an average normalized density ($\rho/\rho_c$) of 
approximately $0.8$.}
\label{fig:denscal}
\end{figure}

Numerical simulations confirm the exponential decay in the
quasi-elastic limit.  Furthermore, the decay length $\xi$ can be
measured for various degrees of inelasticity $r$ 
(see Fig.~\ref{fig:denscal}).  We verified that
indeed $\xi \propto (1-r^2)^{-1/2}$ as suggested by Eq.~(\ref{eqn:hdtemp}). 
The value of the prefactor $\gamma \cong 2.26$ can be
found from the simulations as well.

We now turn to the low density limit.  Here, $z\gg1$, and
Eq.~(\ref{eqn:zgen}) reduces to
$\frac{d^2}{dx^2} z^{3/2} = \frac{16(1-r^2)}{\gamma\alpha^2d^2} z^{-1/2}$.
As a result, 
\begin{equation}
\frac{dz}{dx}=-\frac{1}{\eta}\sqrt{\frac{z-z_L}{z}}
\label{eqn:lddzdx}
\end{equation}
where $\eta \equiv \alpha d \sqrt{3\gamma/64(1-r^2)}$, and 
we have used the boundary condition of Eq.~(\ref{eqn:bcnorm}) 
and the notation $z_L=z(L)$.  Simulations
show that the density near the heated wall is significantly smaller than the
density at the elastic wall.  Thus, for  $x\ll L$ and $z \gg z_L$,
we find the following approximate solution 
$$z(x) \approx z_0 - {x\over\eta}. $$
The inverse of the density, and hence the temperature, decays 
linearly with $x$ in the low density
regime near the heated wall.   That the low density decay length
$\eta\sim (1-r^2)^{-1/2}$ is similar to the high density decay length
$\xi\sim (1-r^2)^{-1/2}$, reflects the fact that the underlying
differential equation is second order.  

\begin{figure}
\narrowtext
\epsfxsize=\hsize
\vspace{-.2in}
\epsfbox{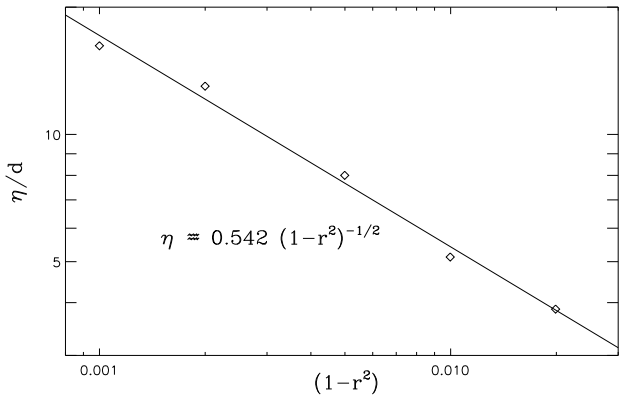}
\vspace{-.2in}
\caption{The behavior of the decay length $\eta$ in units of particle
diameters as a function
of $1-r^2$ in the low density limit.  The solid 
line is $0.542 (1-r^2)^{-1/2}$, and the diamonds are values obtained 
from simulations with an average normalized density ($\rho/\rho_c$) of 
approximately $0.1$.}
\label{fig:dilscal}
\end{figure}

In the nearly elastic case ($1-r \ll 1$), $z_L$ is close to $z$,
and Eq.~(\ref{eqn:lddzdx}) can be integrated exactly
$${x\over \eta}
=- \left[ \sqrt{z}\sqrt{z-z_L} +
 z_L \ln (\sqrt{z}+\sqrt{z-z_L})\right]_{z_0}^{z(x)}.$$
Comparing this prediction with low density numerical simulations 
allows us to determine $\eta$ at various $r$ values 
(see Fig.~\ref{fig:dilscal}) and verify that 
indeed $\eta \propto (1-r^2)^{-1/2}$.  The constant of
proportionality and the previously calculated $\gamma$
yields $\alpha \cong 1.67$.  

In the low density limit, the equation of state $P = \rho T$, 
implies that $T(x) = \frac{P}{\rho_c}z(x)$.  For the state to be
locally very close to equilibrium, the mean
free path must be much less than the length scale over which the 
temperature is changing, i.e., $l \ll \eta \rho_c /P$.  This 
condition is
\begin{equation}
\sqrt{1-r^2}\ll \alpha \sqrt{\frac{\gamma}{2}},
\label{eq:valid2}
\end{equation}
when the temperature is of order unity.  Since
$\alpha$ and $\gamma$ are also of order unity, the hydrodynamic
description is again valid only in the quasi-elastic 
regime ($\sqrt{1-r} \ll 1$).

\begin{figure}
\narrowtext
\epsfxsize=\hsize
\vspace{-.1in}
\epsfbox{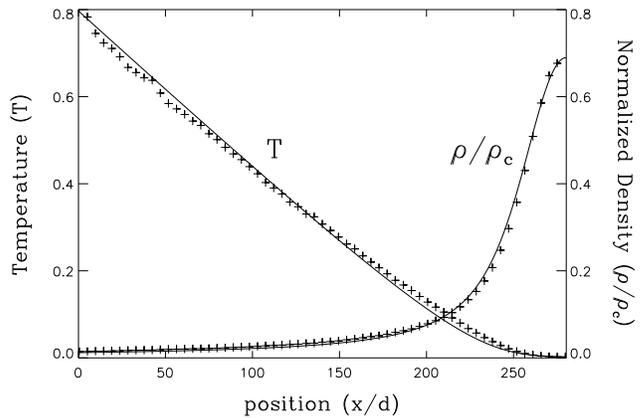}
\vspace{-.1in}
\caption{Comparison of a simulation (plus signs) of $1000$ particles
at $r=0.99$ with the numerical solution (solid lines) to the full differential 
equation~(\ref{eqn:zgen}) with $\gamma=2.26$ and $\alpha=1.15$. The
temperature is not one at the heated wall due to the effect discussed
in Section III.  The horizontal axis is the distance from the heated wall 
in units of the particle diameter ($d$).}  
\label{fig:samp}
\end{figure}

Our treatment so far has concentrated on either the high or the low density
limit, where analytical expressions were possible.  For systems that
include both high and low density regions, Eq.~(\ref{eqn:zgen}) can
be solved numerically using the boundary and normalization
conditions~(\ref{eqn:bcnorm}) and the previously calculated values for
$\gamma$ and $\alpha$.  By examining a number of simulations that 
include a range of densities we determined that the optimal values for
the prefactors are $\gamma=2.26$ and $\alpha=1.15$. While this value
of $\gamma$ is consistent with the value obtained in the high 
density calculation, this $\alpha$ value is slightly lower than
our prediction.  A typical system with $r=0.99$ is shown in
Fig.~\ref{fig:samp}, and it is seen that the predictions of 
hydrodynamic theory match the numerical data over wide density 
variations.  For this simulation, the 
ratio $\sqrt{2(1-r^2)/\gamma\alpha^2}$ is approximately $0.1$, which 
is not of order unity, so the condition of Eq.~(\ref{eq:valid2}) 
is satisfied.

\section{Nonequilibrium steady state}

The calculation in the previous section assumes that
the steady state is very close to equilibrium, and that pressure and
temperature can be used to describe the system. This requires 
that condition (\ref{eq:valid1}) is satisfied in the high density limit, 
or condition (\ref{eq:valid2}) is satisfied in the low density limit.

An important question is: when does the behavior of the system changes
qualitatively?   One such transition occurs when $r$ becomes low
enough for the system to undergo inelastic collapse.  Here strong
correlations and large density variations develop, and applying 
hydrodynamics becomes impossible.  Even when $r$ is slightly 
higher than the critical value for inelastic collapse, the
attractors mentioned in the introduction may still be strong 
enough to build correlations.  Where this breakdown occurs
is determined by the degree of inelasticity, the density and the 
total number of particles\cite{mcn96}.  For each pair of
values of density and total number of particles, there is a value of
$r$ which divides two different kinds of behavior: loose sand and
coherent sand.  

However, the theory developed in the previous 
section is strictly for the quasi-elastic limit, $1-r \ll 1$, so 
there may exist systems that, although elastic enough to avoid
inelastic collapse, still have $r$ far enough from one that the 
hydrodynamics do not apply.  Since correlations between particles 
are built up through inelastic collisions, high density regions
are more liable to inelastic collapse\cite{mcn96}, while 
for low density regions,
correlations are harder to establish.  Therefore, we will investigate
the low density limit in order to observe the breakdown of the 
hydrodynamic description as the degree of inelasticity increases.

\begin{figure}
\narrowtext
\vspace{-.15in}
\epsfxsize=\hsize
\epsfbox{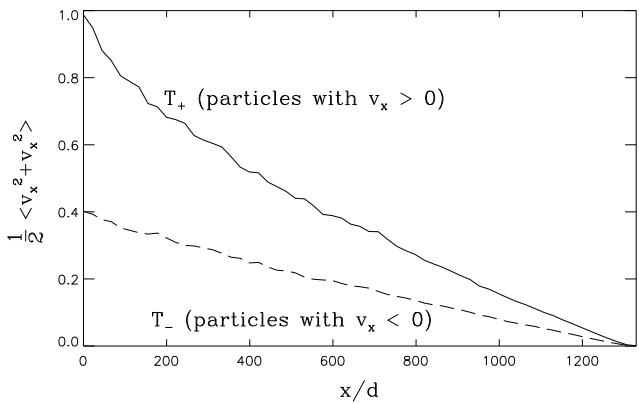}
\vspace{-.15in}
\caption{The difference in temperature for particles with positive 
velocities (solid line) and with negative velocities (dashed line).  
The data is from a simulation 
of 1500 particles with $r=0.95$ and total area fraction $0.01$.}
\label{fig:tempdiff}
\end{figure}

\begin{figure}
\narrowtext
\epsfxsize=\hsize
\vspace{-.15in}
\epsfbox{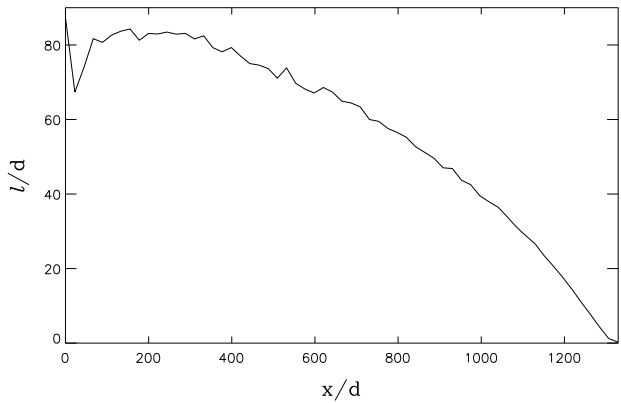}
\vspace{-.15in}
\caption{The mean free path, i.e., the average 
distance until the next collision for 
particles at a given position.  The dip near the heated wall 
is due to collisions with the wall.  The data is from the same
simulation used in Fig.~\ref{fig:tempdiff}.  Both the mean free 
path ($l$) and the position ($x$) are in units of particle 
diameters ($d$).}
\label{fig:mfp}
\end{figure}

When the temperature variation within a mean free path is significant, 
the system is unable to reach local equilibrium.  Therefore, particles
carrying energy away from the heated wall cannot share this energy
with the slow particles returning from the higher density region
near the elastic wall.  This inefficient mixing leads to a temperature
gap -- the average energy of particles with $v_x > 0$ is greater 
than that of the particles with $v_x < 0$ (see Fig.~\ref{fig:tempdiff}).
Furthermore, near the heated wall, the 
temperature drops by approximately $15 \%$ over a
mean free path, which suggests that such a system
will be unable to reach local thermal equilibrium (see Fig.~\ref{fig:mfp}).

\begin{figure}
\narrowtext
\epsfxsize=\hsize
\vspace{-.1in}
\epsfbox{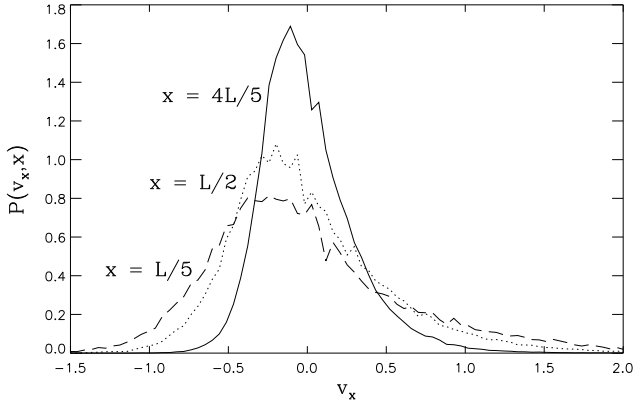}
\vspace{-.1in}
\caption{The probability distributions for $v_x$ at various
positions $x=L/5,L/2,4L/5$, where L is the length of the system 
(1330 particle diameters in this case).  The simulations are for a 
system of $1500$ particles with area fraction $0.01$ at $r=0.95$.}
\label{fig:vxskew}
\end{figure}

\begin{figure}
\narrowtext
\epsfxsize=\hsize
\vspace{-.2in}
\epsfbox{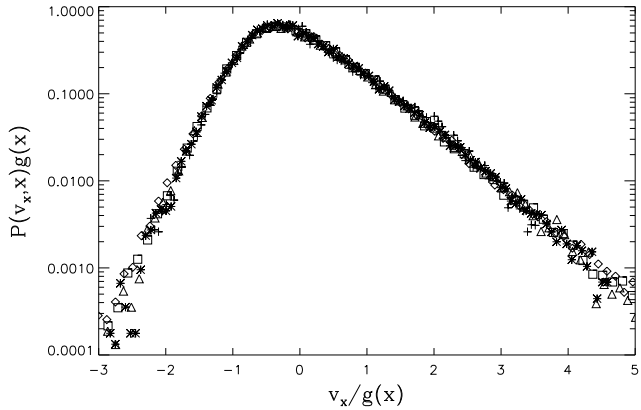}
\vspace{-.1in}
\caption{The scaling distribution $\phi(z) = P(v_x,x)g(x)$.  The
data of Fig.~\ref{fig:vxskew} have been rescaled according to
Eq.~(\ref{eqn:vxprob}), so three sets of data are from various 
positions, $x=L/5(+),L/2(\ast),4L/5(\diamond)$, in a system with 
the usual energy enput at $x=0$.  The plot also includes probability 
distributions from 
a simulation of a system that is identical to that used in 
Fig.~\ref{fig:vxskew} but with a different type of
forcing at the heated wall.  Data from this simulation is displayed 
for $x=7L/10(\triangle),4L/5(\Box)$.}
\label{fig:vxscal}
\end{figure}

The probability distribution functions for $v_x$ (the
velocity perpendicular to the heated wall) provide an
illuminating way to measure this deviation from equilibrium
(see Fig.~\ref{fig:vxskew}).  Note the asymmetry:
the $v_x >0$ tail is longer than the $v_x < 0$ tail.  This is 
consistent with our understanding that the $v_x >0$ particles 
have more energy than the $v_x < 0$ ones.  
A rough quantification of the deviation from equilibrium is 
provided by condition~(\ref{eq:valid2}): 
when $\sqrt{2(1-r^2)/\gamma\alpha^2}$
is of order unity, the theory breaks down.
For $r=0.95$ (the value used in the Figures), this quantity is 
approximately $0.3$.  Note that this condition involves $r$ only
and is not sensitive to the local density $\rho$.  This suggests 
that the behavior of regions of
the system with different densities should be similar.
In fact, we observed that the velocity distribution 
function obeys scaling (see Fig.~\ref{fig:vxscal}), i.e.,
\begin{equation}
P(v_x,x) = \frac{1}{g(x)}\phi\left( \frac{v_x}{g(x)} \right).
\label{eqn:vxprob}
\end{equation}
The function $\phi$ is independent of boundary 
conditions (see Fig.~\ref{fig:vxscal}). 
Thus $\langle v_x^n(x) \rangle \propto g^n(x)$, 
where the constant of proportionality depends only on the
shape of the function $\phi$.  Note that this shape will depend on
$r$, the degree of inelasticity; the more inelastic the
particles are, the more skewed the velocity distribution is.  
The probability distribution for the $y$-component of the 
velocity behaves similarly:  
\begin{equation}
P(v_y,x) = \frac{1}{g(x)}\psi\left( \frac{v_y}{g(x)} \right).
\label{eqn:vyprob}
\end{equation}
but here $\psi$ is a symmetric, nearly Gaussian, function.  
Additionally, the same velocity scale $g(x)$ characterizes 
the transverse and the longitudinal velocity distributions.  
Therefore, while at each position $x$ there is no longer a 
single hydrodynamic temperature, there is a well
defined characteristic velocity scale, $g(x)$, so that the
granular temperature, $\frac{1}{2}\langle v_x^2 + v_y^2\rangle$, is
proportional to $g^2(x)$.  This scaling suggests that,
although the system may deviate significantly from equilibrium,
it can still be treated using some of the tools of the previous
sections.

\begin{figure}
\narrowtext
\epsfxsize=\hsize
\vspace{-.1in}
\epsfbox{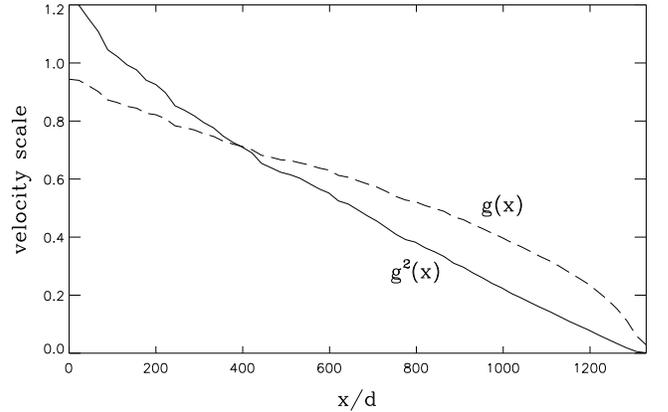}
\vspace{-.1in}
\caption{Characteristic velocity scale as a function of position.
Here we plot $\langle(v_x^2 + v_y^2)^{1/2}\rangle \propto g(x)$
(the dashed line).  This data was obtained from the simulation
used to make Fig.~\ref{fig:vxskew} and the values for $g(x)$ were
used to do the rescaling in Fig.~\ref{fig:vxscal}.  We also
plot $\langle v_x^2 + v_y^2\rangle \propto g^2(x)$ (the solid line)
to show that it is a roughly linear function of position, as
predicted by Eq.~(\ref{eqn:g2linear}).}
\label{fig:gandg2}
\end{figure}

Specifically, return to the equation for energy balance~(\ref{eqn:enerbal}):
$dq/dx = -I$, where $I$ is the energy loss due to collisions
per unit time per unit area, and $q$ is the energy flux,
the heat transfer per unit cross-sectional length per unit time.
As discussed previously,  
$$I \propto (1-r^2) d\rho^2 v^3 \propto \rho^2 g^3,$$
while the heat flux is approximately
$$ q = \langle(\frac{1}{2} \rho v^2) v_x\rangle \propto \rho g^3. $$
Conservation of momentum flux suggests that $\rho g^2$ is constant,  
and the energy balance equation gives
\begin{equation}
\frac{dg}{dx} \propto -g^{-1},
\label{eqn:g2linear}
\end{equation}
which indicates that $g^2$, and hence the temperature, depends linearly
on $x$.  This prediction is consistent with our 
simulational results  (Fig.~\ref{fig:gandg2}).

\section{Conclusion}

In this work, we examined the steady-state behavior of a weakly
inelastic two-dimensional driven granular system.  We found that
a hydrodynamic formulation provides a satisfactory description of the
near-equilibrium behavior of the system in the quasi-elastic limit.
We extended Haff's theory to the low density limit and found that the
corresponding temperature profile varies linearly in space.  For
slightly higher inelasticities, the system is no longer close to
equilibrium in the low density limit. However, the scaling behavior of
the velocity distributions suggests that a hydrodynamic treatment can
still be useful in describing the system.

In this non-equilibrium regime, we found that the particle velocity 
distributions were non-Gaussian.  Indeed,
deviations from normal distributions have been observed in theoretical
\cite{goldzan,lan,tag} and experimental studies\cite{war}.  In addition
to this variation in the velocity distributions, systems of this sort
can produce highly inhomogeneous spatial distributions, as has been noted
elsewhere in one \cite{DLK,GR} and two \cite{EsiPoe,MB} dimensions.  
A recent experimental study examined the spatial distribution of hard 
particles in two dimensions in the presence of an energy input \cite{KWG}. 
Their data is in qualitative agreement with our theoretical predictions, 
and inhomogeneous spatial distributions reminiscent of Fig.~(1) are
observed.

\vspace{.2in}
\noindent {\bf Acknowledgements:}  We are indebted to Leo
Kadanoff for stimulating our interest in this problem.  We
also thank Bill Young, Tom Witten, Sergei Esipov and
Thorsten P\"oschel for useful discussions.
This work was supported by the National Science Foundation
under Awards \#DMR-9415604 and \#DMR-9208527 and the
MRSEC Program, Award \#DMR-9400379.

\end{multicols}

\begin{thebibliography}{99}
\bibitem{JNB}
	H.\ M.\ Jaeger, S.\ R.\ Nagel, and R.\ P.\ Behringer,
	Phys.\ Today  {\bf 49}, 32 (1996).
\bibitem{Haff}
	P.\ K.\ Haff, J.\ Fluid Mech.\ {\bf 134}, 401 (1983).
\bibitem{JenkRich}
	J.\ T.\ Jenkins and M.\ W.\ Richman, J.\ Fluid 
	Mech.\ {\bf 192}, 313 (1988).
\bibitem{DLK}
	Y.\ Du, H.\ Li, and L.\ P.\ Kadanoff, Phys.\ Rev.\ 
	Lett.\ {\bf 74}, 1268 (1995).
\bibitem{GR}
	E.\ L.\ Grossman and B.\ Roman, Phys. Fluids {\bf 8}, 3218 (1996).
\bibitem{Mad}
	J.\ Maddox, Nature {\bf 374}, 11 (1995).
\bibitem{EsiPoe}
	S.\ E.\ Esipov and T.\ P\"{o}schel, ``Boltzmann equation
	and granular hydrodynamics,'' preprint.
\bibitem{hay}
        H.\ Hayakawa, S.\ Yue, and D.\ C.\ Hong, Phys.\ Rev.\ Lett.\ {\bf 75}, 
        2328 (1995).
\bibitem{Tonks}
	L.\ Tonks, Phys.\ Rev.\ {\bf 50}, 955 (1936).
\bibitem{goldzan} I.\ Goldhirsch and G.\ Zanetti, Phys.\ Rev.\ Lett.\
	{\bf 70}, 1619 (1993).
\bibitem{mcn96}
	S.\ McNamara and W.\ R.\ Young, Phys.\ Rev.\ E {\bf 53}, 5089 (1996).
\bibitem{lan}
        Y.\ D.\ Lan and A.\ D.\ Rosato, Phys.\ Fluids {\bf 7}, 1818 (1995).
\bibitem{tag}
        Y.\ H.\ Taguchi and H.\ Takayasu, Europhys.\ Lett.\ 
        {\bf 30}, 499 (1995).
\bibitem{war}
        S.\ Warr, G.\ T.\ H.\ Jaques, and J.\ M.\ Huntley, 
        Powder Tech.\ {\bf 81}, 41 (1994). 
\bibitem{MB}
	S.\ McNamara and J.-L.\ Barrat, ``The energy flux into a
	fluidized granular medium at a vibrating wall,'' preprint.
\bibitem{KWG}
	A.\ Kudrolli, M.\ Wolpert, and J.\ P.\ Gollub, ``Cluster 
	formation due to collisions in granular material,'' preprint.


\end{thebibliography}
\end{document}